\documentclass{aa} 

\usepackage{natbib}
\bibpunct{(}{)}{;}{a}{}{,} 

\usepackage{graphicx}

\usepackage{amssymb,amsmath}
\usepackage{delarray}
\usepackage{mathrsfs}
\usepackage{xcolor}
\usepackage[varg]{txfonts}
\usepackage{hyperref}
\hypersetup{
  colorlinks   = true, 
  urlcolor     = blue, 
  linkcolor    = blue, 
  citecolor   = blue 
}

\begin{document} 


\title{Discontinuities in numerical radiative transfer}

\author{Gioele Janett\inst{1,}\inst{2}
       }
\institute{Istituto Ricerche Solari Locarno (IRSOL), 6605 Locarno-Monti, Switzerland
          \and
          Seminar for Applied Mathematics (SAM) ETHZ, 8093 Zurich, Switzerland
\\          \email{gioele.janett@irsol.ch
          }
          }
 
\abstract
{Observations and magnetohydrodynamic simulations of solar and stellar atmospheres reveal an intermittent behavior or steep gradients 
in physical parameters,
such as magnetic field,
temperature, and bulk velocities.
The numerical solution of the stationary radiative transfer equation is particularly challenging in such situations,
because standard numerical methods may perform very inefficiently in the absence of local smoothness.
However, a rigorous investigation of the numerical treatment of the radiative transfer equation in discontinuous media
is still lacking.
The aim of this work is to expose the limitations of standard convergence analyses for this problem and to identify the relevant issues.
Moreover, specific numerical tests are performed. These show that discontinuities in the atmospheric physical parameters
effectively induce first-order discontinuities in the radiative transfer equation, 
reducing the accuracy of the solution and thwarting high-order convergence.
In addition, a survey of the existing numerical schemes for discontinuous ordinary differential systems
and interpolation techniques for discontinuous discrete data is given,
evaluating their applicability to the radiative transfer problem.
}


\maketitle

\section{Introduction}\label{sec:sec1}
%
The presence of sharp variations in the physical parameters of solar and stellar atmospheres is anything but irrelevant in the formation
of spectroscopic and spectropolarimetric signals.
For instance, it is well known that the propagation of shock waves can generate a large variety of important features in observed spectra \citep{mihalas1980}.
Moreover, the presence of transition layers or boundaries associated with a jump in magnetic
field, in temperature, and in bulk velocities significantly modifies spectral line profiles.

In many astrophysical applications, one must routinely solve the radiative transfer equation.
Very common examples are: iterative solutions of nonlinear radiative
transfer problems in non-local thermodynamic equilibrium (non-LTE) conditions,
inversions of spectral line profiles, and massive three-dimensional (3D) radiative magnetohydrodynamic simulations (R-MHD) of stellar atmospheres.

The radiative transfer equation seldom has
solutions that can be expressed in an analytical form, and therefore it is common to seek approximate
solutions in terms of numerical schemes.
However, standard numerical methods 
rely on smoothness assumptions regarding the functions to be integrated
and these methods may become inaccurate or inefficient in the presence of discontinuities.
In particular, the crossing of a discontinuity might introduce important local errors in the numerical solution,
which negatively affect the accuracy of the global problem.

The relevance of discontinuities has already been recognized in the radiative transfer community.
For example, standard non-LTE radiative transfer codes barely converge
when applied to data from 3D R-MHD simulations that contain discontinuities and steep gradients \citep{steiner2016}.
Some efforts have already been exercised to handle such discontinuities:
\citet{tscharnuter1977} proposed a numerical method for computing the radiative transfer across a shock front with spherical symmetry.
Afterwards, \citet{mihalas1980} designed a numerical scheme for solving the equation of transfer in discontinuous media.
\citet{auer+paletou1994} remarked the necessity to avoid negative overshoots when interpolating physical quantities.
For this reason, \citet{auer2003} suggested the use of Hermite interpolants and Bézier curves in the formal
solution and, subsequently, \citet{ibgui2013} made use of monotonic cubic Hermite polynomials.

Some attention to discontinuities has also been paid in the context of the radiative transfer of polarized light.
For example, \citet{steiner2000} and \citet{mueller2002} studied the role played by discontinuities
in fluid velocities and in the magnetic field in the formation of asymmetric Stokes profiles.
The SIRJUMP inversion code by \citet{louis2009} is able to infer possible discontinuities
in the physical atmospheric quantities along the line of sight \citep{deltoro_iniesta2016}. More recently,
\citet{delacruz_rodriguez+piskunov2013} and~\citet{stepan+trujillo_bueno2013} applied B\'ezier interpolants to control abrupt changes in the absorption and emission quantities, whereas
\citet{steiner2016} proposed a numerical scheme based on monotonic reconstruction that allows for discontinuities at the boundary of each numerical cell of the atmospheric model.

However, a rigorous mathematical and numerical investigation of the problems arising in the integration of the discontinuous radiative transfer equation
is still lacking and deeper understanding of this topic is certainly required. 
This work clearly states the issues introduced by discontinuities in the context of the radiative transfer equation.
The paper is organized as follows: 
Sections~\ref{sec:sec2} and~\ref{sec:sec31} introduce differential systems with a discontinuous right-hand side and 
interpolations of discontinuous functions, respectively,
explaining the limitations of standard convergence analyses for these kinds of problems.
Section~\ref{sec:sec3} explores the role of discontinuities in the radiative transfer equation. 
Section~\ref{sec:sec5} exposes specific numerical tests that highlight
the inefficient performances of standard methods when dealing with discontinuities in the radiative transfer equation.
Sections~\ref{sec:sec4} and~\ref{sec:int_disc_data} briefly summarize the existing numerical methods for the treatment of discontinuous ordinary differential equations (ODEs)
and the interpolation techniques for discontinuous discrete data.
Particular attention is paid to their suitability to the radiative transfer equation in discontinuous media.
Finally, Section~\ref{sec:sec6} provides remarks and conclusions, with a view on future work.
\section{Ordinary differential
equations with a discontinuous right-hand side}\label{sec:sec2}
Discontinuous differential systems frequently appear in various fields, such as chemical kinetics \citep[e.g.,][]{landry2009}, biological systems \citep[e.g.,][]{gouze2002}, and various engineering disciplines \citep[e.g.,][]{malmborg1997,acary2008}.
Moreover, most of the studies on discontinuous ODEs rely on numerical methods and a broad literature attests to the importance of this topic:
from the pioneering works by \citet{chartres1972} and \citet{mannshardt1978} to the surveys by \citet{calvo2008} and \citet{dieci2012}.
\subsection{Impact of a discontinuity on the numerical solution}\label{sec:sec2_3}
The numerical integration of discontinuous differential systems is challenging, because standard numerical schemes may perform very inefficiently in the absence of local smoothness.
It is known that the crossing of a discontinuity introduces significant local errors, which negatively affect global errors 
and might even yield a convergence order breakdown.

Consider a numerical method of order $p$ applied to the scalar initial value problem (IVP)
\begin{equation}\label{scalarIVP}
y'(t) = f(t,y)\,, \quad y(0) = y_0\,.
\end{equation}
Standard convergence analysis would conclude that the numerical scheme yields a global error that scales as $\mathcal{O}(h^p)$, where $h$ denotes the step size.
Such an analysis relies on the assumption that
the right-hand side $f$ of the IVP~\eqref{scalarIVP} (and consequently $y$) is sufficiently smooth.
More precisely, the function $f$ must have $p+1$ continuous derivatives.
However, a different local truncation error analysis is mandatory if discontinuities occur in $f$ or
its derivatives.
By definition, a discontinuity in Equation~\eqref{scalarIVP} is of order $q \ge 1$ if $f$ contains a finite jump
in at least one of its partial derivatives of order $q - 1$ and
it has continuous derivatives through order $q - 2$ \citep{gear1984}.
This produces a $q$th-order discontinuity in $y$ at the discontinuity location.

Numerical analysis shows that if $q \ge p + 1$ the numerical method will remain of order $p$ and one may not even notice the discontinuity. 
By contrast, if $q \le p$, the discontinuity introduces the dominant term $\mathcal{O}(h^q)$ in the local error, which can reduce the global error order.
For instance, \citet{mannshardt1978} observed that a Runge-Kutta method remains convergent after having crossed a first-order discontinuity, but only with order 1.

\begin{figure}
\includegraphics[width=0.49\textwidth]{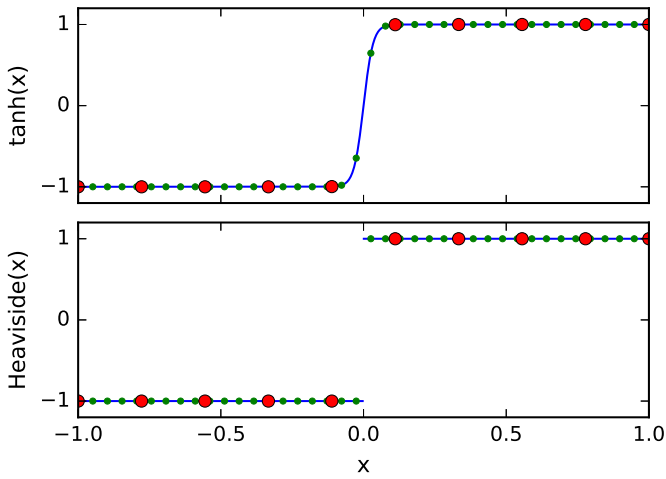}
  \caption{The hyperbolic tangent (top panel, continuous blue) and the modified Heaviside (bottom panel, continuous blue)
  functions are sampled in the interval $x\in[-1,1]$ with 10 (red circles) and 40 (green dots) grid points.
  No difference between the coarser (red) samplings is visible, while this is not true for the finer (green) samplings.}
  \label{fig:sampling}
\end{figure}

In many practical problems, the right-hand side $f$ of the IVP~\eqref{scalarIVP} is known at
a discrete set of points only. In such cases, the numerical grid must be considered fixed \textit{a priori}
and there is not a well defined concept of discontinuity in $f$.
It is difficult to tell whether the underlying model that gave
rise to the discrete data is continuous, or rather contains jumps.
In a discrete world, all changes are jumps by definition
and data are purely discontinuous even if the underlying model is not.
Therefore, there is no concrete difference between a discontinuity and a steep gradient
for a single set of discrete data that is not able to resolve the local feature as shown in Figure~\ref{fig:sampling}.
The difference is only observed if one considers
a sequence of discrete grids with a finer and finer sampling.
Computing the maximum of the differences between adjacent data points along the sequence of increasingly refined grids, one recognizes two possible cases:
either the underlying function is continuous and the value approaches zero,
or it contains a discontinuity and the value approaches the size of the jump.

Practical radiative transfer problems often involve rapidly varying functions which actually do not contain any mathematical discontinuities.
Section~\ref{sec:sec5.1} numerically explores such issues by considering both a discontinuous atmospheric model
and one containing a steep gradient.

\subsection{Example: trapezoidal method order reduction}\label{sec:subsec2.2}
This section gives a concrete example of order reduction for the (implicit) trapezoidal method,
but the following local error analysis can be adapted to other numerical methods.
For instance, \citet{mannshardt1978} presented the case of a general one-step method, whereas \citet{gear1984} analyzed predictor-corrector methods.

For simplicity, consider the IVP
\begin{equation}\label{scalarIVP_disc}
y'(t) = f(t)=
\begin{cases}
f_1(t) & \text{if}\; t<\xi\,, \\ 
f_2(t) & \text{if}\; t\ge\xi\,,
\end{cases}
\quad y(0) = y_0\,,
\end{equation}
where $f:\mathbb{R}\rightarrow\mathbb{R}$
and the discretization of the time interval $[0,T]$ given by $\{t_k\}$ with $k=0,\dots,N$.
Let $\xi\in[t_k,t_{k+1}]$ and assume both $f_1$ and $f_2$ at least three-times continuously differentiable. 
For $f_1(\xi)\ne f_2(\xi)$, the IVP~\eqref{scalarIVP_disc} shows a first-order discontinuity.
The trapezoidal method applied to the IVP~\eqref{scalarIVP_disc} in the interval $[t_k,t_{k+1}]$ reads
\begin{equation}
y_{k+1} = y_k+\frac{t_{k+1}-t_k}{2}\left[f_1(t_k)+f_2(t_{k+1})\right]\,,
\label{trapezoidal}
\end{equation}
where $y_k$ and $y_{k+1}$ are the numerical approximations of the exact values $y(t_k)$ and $y(t_{k+1})$, respectively. 
The local truncation error is defined by
\begin{equation*}
L_{k+1}=y(t_{k+1})-y_{k+1}\,,\text{ assuming }y_{k}=y(t_{k})\,,
\end{equation*}
while the global error, defined as
\begin{equation*}
E_N= |y_N-y(t_N)|\,,
\end{equation*}
represents the accumulation of local errors over all the steps.
The Taylor expansion of the local exact solution gives
{\small
\begin{align*}
y(t_{k+1}) &= y(\xi)+(t_{k+1}-\xi)f_2(\xi)+\frac{(t_{k+1}-\xi)^2}{2}f_2'(\xi)+\mathcal{O}((t_{k+1}-\xi)^3)\,,\\
y(\xi) &= y_k+(\xi-t_k)f_1(t_k)+\frac{(\xi-t_k)^2}{2}f_1'(t_k)+\mathcal{O}((\xi-t_k)^3)\,.
\end{align*}}\noindent
Making use of the additional Taylor expansions
\begin{align*}
f_1(t_k) &= f_1(\xi)+(t_k-\xi)f_1'(\xi)+\mathcal{O}((\xi-t_k)^2)\,,\\
f_1'(t_k) &= f_1'(\xi)+\mathcal{O}(t_k-\xi)\,,\\
f_2(t_{k+1}) &= f_2(\xi)+(t_{k+1}-\xi)f_2'(\xi)+\mathcal{O}((t_{k+1}-\xi)^2)\,,
\end{align*}
one recovers by direct calculations the expression
\begin{equation*}
\begin{split}
L_{k+1}&=(\xi-\frac{t_k+t_{k+1}}{2})\left[f_1(\xi)-f_2(\xi)\right]\\
&+(\xi-t_k)(t_{k+1}-\xi)\left[f_1'(\xi)-f_2'(\xi)\right]+\mathcal{O}(h^3)\,,
\end{split}
\end{equation*}
where $h=t_{k+1}-t_k$.
This leads to the formula
\begin{equation}
|L_{k+1}|\approx hK_1+h^2K_2+ \mathcal{O}(h^3)\,,
\label{local_err_disc}
\end{equation}
where
\begin{align*}
K_1&=|f_1(\xi)-f_2(\xi)|\,,\\
K_2&=|f_1'(\xi)-f_2'(\xi)|\,,
\end{align*}
are the first and second-order jumps, respectively, at the discontinuity point $\xi$.
The first and second terms in Equation~\eqref{local_err_disc} contribute to the global error 
with $\mathcal{O}(h)$ and $\mathcal{O}(h^2)$ terms, respectively.
These contributions arise from the jump and cannot be improved by using some higher-order method.
After having crossed a first-order discontinuity, that is, $f_1(\xi)\neq f_2(\xi)$,
the trapezoidal method is therefore only first-order convergent. Here, one assumes a finite number of jumps,
otherwise the $\mathcal{O}(h)$ contributions will accumulate, thwarting convergence.
Discontinuities of second order or greater (which imply $f_1(\xi)=f_2(\xi)$) do not affect the order of convergence of the trapezoidal method.
\section{Interpolations of discontinuous functions}\label{sec:sec31}
Consider a function $y(x)$ and a discrete set of points $x_0< x_1<\ldots< x_n$ and
assume that $y(x_0),\dots, y(x_n)$ are given.
The interpolation problem is to find a function $p(x)$ that satisfies
interpolation requirements
\begin{equation}\label{interp_cond}
p(x_i)=y(x_i)\, \text{ for } i=0,\dots,n\,.
\end{equation}
The simplest way is to connect the
given points with straight lines. However, when seeking for higher accuracy, it is usual to look for a function $p(x)$
that is a polynomial or a piecewise-smooth polynomial of higher degree.
Alternatively, the function $p(x)$
can be a trigonometric function, a rational polynomial function, and so on.
\subsection{Polynomial interpolation}\label{pol_int}
In the problem of polynomial interpolation, one seeks a polynomial $p(x)$ that
satisfies the interpolation condition~\eqref{interp_cond}.
Limiting the degree of
$p(x)$ to be $\le n$, one obtains precisely one single interpolant $p_n(x)$ that satisfies the interpolation conditions.
A very common strategy is to construct this interpolant in terms of Lagrange polynomials, namely
\begin{equation}\label{lagrange_interp}
p_n(x)=\sum_{i=0}^{n}y(x_i)\ell_i(x)\,,
\end{equation}
%
where the Lagrange basis polynomials, $\ell_i(x)$, given by
\begin{equation*}
\ell_i(x)=\prod_{\substack{0\le m \le n\\m\neq i}}\frac{x-x_m}{x_i-x_m}\,,
\end{equation*}
satisfy the relation $\ell_i(x_j)=\delta_{ij}$, with $\delta_{ij}$  being the Kronecker delta
\begin{equation*}
\delta_{ij} = 
\begin{cases}
1 & \text{ if } i = j \,, \\ 
0 & \text{ if } i \ne j\,.
\end{cases}
\end{equation*}
The interpolant $p_n(x)$ and the interpolated function $y(x)$ in Equation~\eqref{lagrange_interp}
satisfy the condition~\eqref{interp_cond},
that is, they agree with each other at the interpolation points.
In general, there is no reason to expect them to be close to each other elsewhere. Nevertheless, one can estimate
the difference between them, the so-called interpolation error.

Let $\Pi_n$ denote the space of polynomials of degree $\le n$, and let $C^{n+1}[a,b]$ denote the
space of functions that have $n+1$ continuous derivatives on the interval $[a, b]$.
Suppose $y(x)\in C^{n+1}[a,b]$ and let $p_n(x)\in \Pi_n$ be the unique polynomial
that interpolates $y(x)$
at the $n+1$ distinct points $x_0< x_1<\dots< x_n \in [a,b]$.
Then, for all $x \in [a,b]$ there is a $\xi \in (a,b)$, such that
\begin{equation}\label{pol_accuracy1}
y(x)-p_n(x) =\frac{1}{(n+1)!}y^{(n+1)}(\xi)\prod_{j=0}^n(x-x_j)\,.
\end{equation}
Defining $h= \max_j x_{j+1}-x_j$, one obtains
\begin{equation}\label{pol_accuracy2}
\vert y-p_n\vert_\infty \coloneqq   \sup_{a\le x\le b}|y(x)-p_n(x)|\le \frac{h^{n+1}}{4n}\vert y^{(n+1)} \vert_\infty\,. 
\end{equation}
This means that, by assuming $y(x)$ smooth enough,
the error decreases to zero as $\mathcal{O}(h^{n+1})$ and the interpolation has an order of accuracy $n+1$.
However, the smoothness assumption of $y(x)$ is not always fulfilled
and polynomial interpolations may become inaccurate in the presence of discontinuities.

Moreover, the error of the interpolation does not necessarily decrease by increasing
the order of the polynomial.
In fact, even though
$h^{n+1}$ may go to zero for $n\rightarrow\infty$, the term $|y^{(n+1)}|$ can grow rapidly, preventing convergence in Equation~\eqref{pol_accuracy2}.
Equidistant interpolation of the Runge’s function (the so-called Runge phenomenon) is a striking example of this.
%
\subsection{Impact of discontinuities}\label{subsec_int_disc}
Error estimates~\eqref{pol_accuracy1} and~\eqref{pol_accuracy2} assume some smoothness of $y(x)$
and the order of accuracy of interpolations decreases for less regular functions.
Suppose that $y(x) \in C^{s}[a,b]$ with $s<n$.
Then, Equation~\eqref{pol_accuracy1} must be replaced by
\begin{equation}
y(x)-p_n(x)=\frac{y^{(s)}(\xi)-p_n^{(s)}(\xi)}{n!}(x-\tilde x_1)(x-\tilde x_2)\dots(x-\tilde x_s)\,,
\label{disc_interp_err}
\end{equation}
where $\{\tilde x_j\}^s_{j=1}$ is any subset of $\{x_j\}^{n}_{j=0}$.
By the same arguments that led up to Equation~\eqref{pol_accuracy2},
one obtains 
\begin{equation*}
\vert y-p_n\vert_\infty = \mathcal{O}(h^{s})\,, 
\end{equation*}
that is, the order of accuracy of the polynomial interpolation is reduced to $s$.

Moreover, it is known that standard second-order or higher interpolations are oscillatory near discontinuities and such oscillations might
lead to numerical inaccuracy and even to numerical instability in nonlinear problems \citep[e.g.,][]{richards1991,shu1998}.
This implies, first, an over/undershoot in $p_n(x)$ ($n>1$) whenever the function $y(x)$ has a jump discontinuity
and, second, the failure of higher-order approximations to remove the overshoot.
This behavior (similar to the Gibbs phenomenon in Fourier series) is known as Gibbs phenomenon.
For example, \citet{zhang1997} showed that cubic spline interpolations on uniform meshes always oscillate near a jump discontinuity.
For these reasons, it makes little sense (and could even be noxious) to use standard high-order interpolation $(n>s)$ when the smoothness of $y(x)$ is not guaranteed.
\subsection{Interpolatory quadratures}\label{subsec_int_quad}
Consider a function $y(x)$ and assume that a discrete set of points $x_0< x_1<\ldots< x_n$ and the data values
$y(x_0),\dots, y(x_n)$ are given.
A systematic approximation of the integration of $y(x)$ in an interval $[\tilde a,\tilde b]$ is commonly called quadrature
and its formula is usually given in the form
\begin{equation}\label{quadrature}
\int_{\tilde a}^{\tilde b} y(x){\rm d}x\approx\sum_{i=0}^n y(x_i)\omega_i\,,
\end{equation}
where $\omega_i$ are the quadrature weights.
A usual way to derive a quadrature formula is through the so-called interpolatory quadrature:
one makes an assumption for the polynomial form of the integrand in the integration interval and analytically solves the integral.
For instance, the Lagrange interpolation given by Equation~\eqref{lagrange_interp} yields
\begin{equation*}
\int_{\tilde a}^{\tilde b} y(x){\rm d}x\approx \sum_{i=0}^n y(x_i)\int_{\tilde a}^{\tilde b} \ell_i(x){\rm d}x\,,
\end{equation*}
where the integrals of the Lagrange basis polynomials provide the quadrature weights.
The quadrature error, i.e., the difference between the left-hand and the right-hand sides
of Equation~\eqref{quadrature}, decreases to zero as $\mathcal{O}(h^{n+2})$.
Moreover, the quadrature is exact for all polynomials of degree $\le n$, that is, for all $y(x)\in\Pi_{n}$.

However, Section~\ref{subsec_int_disc} shows that a polynomial of degree $n$ guarantees $\mathcal{O}(h^{n+1})$ accuracy only if
$y(x) \in C^{n+1}$ in the integration interval. 
Consequently, interpolatory quadratures suffer from the same order-reduction issue of interpolations when facing discontinuities.
Moreover, the presence of overshoots in the interpolated integrand
reduces the accuracy of the quadrature
and could even yield negative results in the quadrature of positive integrands.
For this reason, the use of standard high-order interpolatory quadratures is not encouraged when the smoothness of $y(x)$ is not guaranteed.
\section{Discontinuities in radiative transfer}\label{sec:sec3}
Standard radiative transfer calculations usually assume smooth variations in the radiation field and in the atmospheric physical parameters.
The incidence of discontinuities is often neglected, without considering 
whether this assumption is justified or not.
This section inquires into the role of discontinuities in the radiative transfer equation,
focusing on the possible numerical issues.
\subsection{Discontinuities in solar and stellar atmospheres}
The plasma of solar and stellar atmospheres can be highly dynamic,
with its properties fluctuating sharply over small scales.
For instance, jumps in the magnetic field frequently appear: 
a typical example is the magnetopause at the boundary of a flux tube, where
the magnetic field of the tube is separated from the surrounding field-free
plasma by a thin transition layer.
Another example is given by magnetic canopies, that is, horizontally
extending magnetic fields overlying field-free regions of internetwork cells and
separated from it by a thin transition layer \citep{steiner2000}.
Abrupt variations in temperature can also be present:
for example, the jump in temperature of three orders of magnitude that takes place in the solar transition region.
Clouds of cool material embedded in the chromosphere or corona
are usually associated with jumps in bulk velocities as well.
For instance, a magnetic flux tube interacting with convective motions shows a downwardly directed flow along its boundary \citep{steiner1998}.
The presence of shock fronts could also imply discontinuities in the velocity field.
All these sudden variations of the physical quantities describing the atmosphere might reflect into sudden variations in absorption and emissivity. 

Radiative transfer calculations make use of discrete atmospheric models and
one can distinguish between two cases:
on the one hand, there are semi-empirical atmospheric models \citep[e.g.,][]{vernazza1981,fontenla1990,fontenla1993}, which are still frequently used to synthesize spectral lines.
In these models, the physical parameters across the atmosphere usually vary smoothly.
On the other hand, state-of-the-art 3D R-MHD simulations of stellar atmospheres show contact discontinuities, steep
gradients in state variables, and shock fronts.
In fact, if the MHD equations are solved by the method of characteristics
then shocks are represented by mathematical discontinuities \citep{mihalas1980},
while in the case of a nonideal fluid, sharp structures might be smeared over a few zones in the mesh.
Nonetheless, one still deals with a quasi discontinuity when the jump in physical states occurs
on a spatial scale much smaller than that of the
numerical discretization \citep{steiner2016}.

Discrete models provide the physical parameters describing the atmosphere at a discrete set of spatial points only.
Section~\ref{sec:sec2_3} explains that, in this case,
there is no well defined concept of discontinuity.
However, one is still interested in determining the possible impact of such features on the numerical solution.
\subsection{Radiative transfer equation}
The monochromatic (time-dependent) transfer of unpolarized 
light is described by the following scalar partial differential equation \citep{mihalas1978}
\begin{equation}
  \left[\frac{1}{c}\frac{\partial}{\partial t}+\frac{\partial}{\partial s}\right] I_{\nu}(s,t)  = -\chi_{\nu}(s,t) I_{\nu}(s,t) + \epsilon_{\nu}(s,t)\,,
\label{eq:general_scalar_RTE}
\end{equation}
where $I_{\nu}$ is the specific intensity of radiation propagating in the ray path direction at time $t$. Moreover, $\chi_{\nu}$ and $\epsilon_{\nu}$ are the absorption and the emission coefficients, respectively, $\nu$ is the frequency, and $c$ is the speed of light. The spatial coordinate $s$ denotes the position along the ray under consideration. Equation~\eqref{eq:general_scalar_RTE} is a hyperbolic partial differential equation and, specifically, an advection equation with source and sink terms.
It is known that computing the numerical solution of advection equations can be particularly challenging.
However, in the vast majority of radiative transfer problems in astrophysics, the photon free-flight time 
is much smaller than the fluid dynamics timescales and the radiation field fully stabilizes before any change occurs in the medium.
Therefore, one usually assumes the steady-state version of Equation~\eqref{eq:general_scalar_RTE}, 
consisting in the linear first-order inhomogeneous ODE given by \citep{mihalas1978}
\begin{equation}
  \frac{\rm d}{{\rm d} s} I_{\nu}(s) = -\chi_{\nu}(s) I_{\nu}(s) + \epsilon_{\nu}(s)\coloneqq F_{\nu}(s,I_{\nu}(s))\,.
  \label{eq:scalar_RTE}
\end{equation}
Assuming that $F_{\nu}$ is Riemann integrable in the interval $[s_0,s]$, one formally integrates Equation~\eqref{eq:scalar_RTE} and obtains
\begin{equation}
I_{\nu}(s)=I_{\nu}(s_{0})+\int_{s_{0}}^{s}F_{\nu}(x,I_{\nu}(x)){\rm d}x\,.
\label{scalar_solution}
\end{equation}
Here, the problem of calculating the emergent radiation consists in the evaluation of the integral in the right-hand side of Equation~\eqref{scalar_solution},
which in turn depends on the specific intensity $I_{\nu}$.

When dealing with radiative transfer in astrophysical plasmas, it is common to refer to the concept of optical depth. Replacing the coordinate $s$ by the optical depth $\tau_{\nu}$ defined by
\begin{equation}
{\rm d}\tau_{\nu}=-\chi_{\nu}(s){\rm d}s\,,
\label{opt_depth}
\end{equation}
one recasts Equation~\eqref{eq:scalar_RTE} into
\begin{equation}
  \frac{\rm d}{{\rm d} \tau_{\nu}} I_{\nu}(\tau_{\nu}) = I_{\nu}(\tau_{\nu}) - S_{\nu}(\tau_{\nu})\,,
\label{eq:scalar_RTE_tau}
\end{equation}
where $S_{\nu} = \epsilon_{\nu}/\chi_{\nu}$
is the source function, that is, the ratio between the emission and absorption coefficients.
Moreover, the formal integration of Equation~\eqref{eq:scalar_RTE_tau} in the interval $\left[\tau_{\nu,0},\tau_{\nu}\right]$
gives
\begin{equation}
I_{\nu}(\tau_{\nu})=e^{-(\tau_{\nu,0}-\tau_{\nu})}I_{\nu}(\tau_{\nu,0})+\int^{\tau_{\nu,0}}_{\tau_{\nu}}e^{-(x-\tau_{\nu})} S_{\nu}(x){\rm d}x\,,
\label{scalar_solution_tau}
\end{equation}
where $\tau_{\nu,0}\ge\tau_{\nu}$. Here, the problem of calculating the emergent radiation
reduces to an integral evaluation.
%
\subsection{Numerical approximations}
\citet{auer2003} defined the (numerical) formal solution of Equation~\eqref{eq:scalar_RTE} as the evaluation of the radiation field,
given knowledge of the opacity $\chi_k$ and the emissivity $\epsilon_k$
coefficients at each grid point $s_{k}$, and of the boundary conditions.

The first step in the numerical approach to an ODE involves the discretization of the integration domain. 
Therefore, one discretizes the ray path under consideration with a spatial depth grid $\{s_k\}$ (or an optical depth grid $\{\tau_k\}$) with $k=0,\dots,N$.
The index $k$ increases along the propagation direction. The numerical approximation of a certain quantity at node  $s_k$ (or $\tau_k$)
is indicated by substituting the explicit dependence on $s$ (or $\tau$) with the subscript $k$, for instance,
\begin{equation*}
I_k \approx I(s_k)\,.
\end{equation*}
For notational simplicity, the frequency dependence is omitted.
Inserting numerical quantities in Equation~\eqref{scalar_solution}, one obtains\footnote{The very same procedure can apply starting from
Equation~\eqref{eq:scalar_RTE_tau} instead of Equation~\eqref{eq:scalar_RTE}. \citet{janett2018a} analyze the differences
between the two cases.}
\begin{equation*}
I_{k+1}=I_k+\int_{s_k}^{s_{k+1}}F(s,I(s)){\rm d}s\,.
\end{equation*}
Different approximations of the integral on the right-hand side yield different numerical schemes.
Well-known examples are the backward Euler method, the implicit trapezoidal method,
and the third-order Runge-Kutta method (see Table~\ref{tab:schemes}).
Note that some numerical schemes (e.g., Runge-Kutta 3) use intermediate grid points for the integration.
In this case, one must recover the relevant quantities at off-grid points, which are typically provided through interpolation.
This procedure may alter absorption and emission coefficients and introduce numerical errors.


%
%

However, the most common formal solutions
ground on Equation~\eqref{scalar_solution_tau}.
Inserting numerical quantities, one gets 
\begin{equation}
I_{k+1}=e^{-\Delta\tau_k}I_k+\int^{\tau_k}_{\tau_{k+1}}e^{-(x-\tau_{k+1})} S(x){\rm d}x\,,
\label{sc_formal_solution}
\end{equation}
where,
from Equation~\eqref{opt_depth}, one has
\begin{equation}
\Delta \tau_k = \tau_k -\tau_{k+1} = \int_{s_k}^{s_{k+1}}\chi(s){\rm d}s\,.
\label{conversion_opt_depth}
\end{equation}
This recursive relation is the starting point for 
the so-called exponential integrators \citep{hochbruck2010}, which are a family of numerical schemes
that exhibit strong stability properties\footnote{Positiveness of $\Delta \tau_k$ guarantees $L$-stability \citep{janett2018a}.}.
The numerical problem is reduced to the quadrature of the integrals in Equations~\eqref{sc_formal_solution} and~\eqref{conversion_opt_depth},
for which different techniques are available.
The knowledge of $S$ and $\chi$ at the neighboring grid points suggests the use of interpolatory quadratures (see Section~\ref{subsec_int_quad}).
The standard choice is to use Lagrange polynomials and the simplest strategy is either constant or linear approximations inside the integration interval.
However, \citet{auer+paletou1994} noted that in order to recover the diffusion approximation, it is necessary to use parabolic or higher-order interpolations of $S$.
\citet*{olson+kunasz1987} and \citet*{kunasz+olson1988} made use of parabolic interpolations and,
alternatively, \citet*{mihalas+auer1978} proposed cubic Hermite interpolations of $S$.

Section~\ref{sec:sec2} explains that
the accuracy of numerical schemes for ODEs rely on different assumptions on the degree of smoothness of the functional $F$.
Moreover, Section~\ref{subsec_int_disc} argued that polynomial interpolations are inaccurate around discontinuities,
whereas Section~\ref{subsec_int_quad} showed that the accuracy of interpolatory quadratures
relies on a sufficient smoothness of the integrand.
\subsection{Polarized radiative transfer equation}\label{section_polarized_rte}
Most classical radiative transfer problems do not consider polarization. However, the transfer of polarized light is of particular interest in many applications. 
The radiative transfer of partially polarized light is described by the system of first-order coupled inhomogeneous ODEs given by
\begin{equation}\label{eq:RTE}
  \frac{\rm d}{{\rm d} s}\mathbf I_{\nu}(s) 
  = -\mathbf K_{\nu}(s)\mathbf I_{\nu}(s) + \boldsymbol{\epsilon}_{\nu}(s)\,,
\end{equation}
where $s$ is the spatial coordinate measured along the ray under consideration, $\mathbf{I}_{\nu}=(I,Q,U,V)^{T}$ is the Stokes vector,
\begin{equation*}
  \mathbf K_{\nu} = \begin{pmatrix}
      \eta_I &  \eta_Q &  \eta_U & \eta_V  \\
      \eta_Q &  \eta_I &  \rho_V & -\rho_U \\
      \eta_U & -\rho_V &  \eta_I & \rho_Q  \\
      \eta_V &  \rho_U & -\rho_Q & \eta_I 
               \end{pmatrix}\,
\label{matrix_K}
\end{equation*}
is the $4\times4$ propagation matrix,
and $\boldsymbol{\epsilon}_{\nu}=(\epsilon_I,\epsilon_Q,\epsilon_U,\epsilon_V)^{T}$ is the emission vector \citep{landi_deglinnocenti+landolfi2004}.
For notational simplicity the frequency dependence of all the vectorial and matricial entries is omitted.

The peculiarity of Equation~\eqref{eq:RTE} originates simply from its matricial character \citep{landi_deglinnocenti+landolfi2004}.
In fact, the assumption of vanishing off-diagonal coefficients in the propagation matrix decouples Equation~\eqref{eq:RTE}
into four independent scalar problems formally identical to Equation~\eqref{eq:scalar_RTE}, reducing the vectorial problem to the scalar one.
The diagonal element of the propagation matrix corresponds to the total absorption coefficient in the unpolarized case,
that is, $\eta_I\equiv\chi_{\nu}$.

The generalization of the definition of the scalar formal solution  to the polarized case consists in substituting radiation intensity, opacity, and emissivity by Stokes vector, propagation matrix, and emission vector, respectively.
The polarized problem is considerably more complex than the scalar one.
For instance, the problem cannot be reduced to simple quadratures \citep[see][]{landi_deglinnocenti+landolfi2004}.
Moreover the radiative transfer equation for polarized light exhibits stiff behavior, and numerical schemes face instability issues \citep{janett2018a}.
An extensive analysis of the numerical solution of the polarized radiative transfer equation is given by \citet{janett2017a,janett2017b},
who characterized several exponential integrators (DELO methods) and 
high-order formal solvers.

Just like in the scalar case, the vectorial formal solution suffers from numerical issues in the presence of discontinuities.
For the sake of simplicity, the full convergence analysis is not repeated,
but the numerical tests presented in Section~\ref{sec:sec5} comprehend the polarized problem.
%
\subsection{Two- and three-dimensional problems}
In 1D problems, any ray under consideration is discretized by the grid points.
However, in 2D or 3D problems,
the ray path (long characteristics) or the downwind and upwind ray segments (short characteristics)
do not generally coincide with the grid points at which the physical quantities are given. 
At each integration step, one must then recover the neighboring absorption and emission (or source function) coefficients
and the upwind intensity $I_k$
at off-grid points\footnote{Multistep methods make also use of additional upwind intensities ($I_{k-1},I_{k-2},\dots)$.
However, it is very uncommon to apply this class of numerical schemes to radiative transfer problems.}.
This is typically performed through interpolations, whose role is to ``fill in the gaps'' between the known discrete values \citep{auer2003,steffen2017}. 

In 2D problems, the usual strategy is to discretize the ray by taking its intersections with the segments connecting the grid points,
dealing then with 1D interpolations to recover the off-grid points quantities \citep{auer2003}.
The linear interpolations from the values of the nearest grid points 
is the simplest and fastest method,
but it leads to a large numerical dispersion \citep{fabianibendicho2003,ibgui2013}.
As the accuracy of the formal solution depends on the accuracy of the interpolated off-grid points
quantities, it is often necessary to use high-order interpolations. 

In 3D problems, an efficient approach is to discretize the ray by taking its intersections with the individual planes defined by the grid points,
dealing then with 2D interpolations.
The simplest approach is to use bilinear interpolations, because a cell wall has four corner points.
Once again, linear interpolations are not suitable to reach high accuracy.  
Extending to higher-order, one can use bi-quadratic or bi-cubic interpolations \citep{dullemond2013}.

Section~\ref{subsec_int_disc} explains that if the behavior of the quantities to be interpolated is discontinuous or particularly intermittent,
high-order interpolations oscillate and may introduce artifacts.
Both \citet{kunasz+auer1988} and \citet{auer2003} highlighted that high-order Lagrange polynomials may introduce spurious extrema
that could lead to nonphysical negative values of the source function or of the intensity.
%
\section{Numerical tests}\label{sec:sec5}
Numerical analysis alone is merely theory. Indeed,
numerical evidence is mandatory to assess the mathematical predictions given in the previous sections.
Numerical evidence for oscillatory interpolations is easily accessible in the literature and therefore is not exposed here.
In order to determine whether and how discontinuities (or high-gradients) affect the numerical  
integration of Equation~\eqref{eq:RTE}, different scenarios are investigated.
Since Equation~\eqref{eq:RTE} is always integrated along a ray \citep{auer2003},
the 1D geometry is adopted.
At first, different analytical models for the scalar radiative transfer are analyzed and,
secondly, the vectorial polarized case is considered.
One should avoid drawing premature conclusions on the mere basis of small details of error curves.
However, the numerical tests presented in this section allow for some general considerations.
Table~\ref{tab:schemes} gives an overview of the numerical schemes used.
\begin{figure*}
\centering
 \includegraphics[width=1.\textwidth]{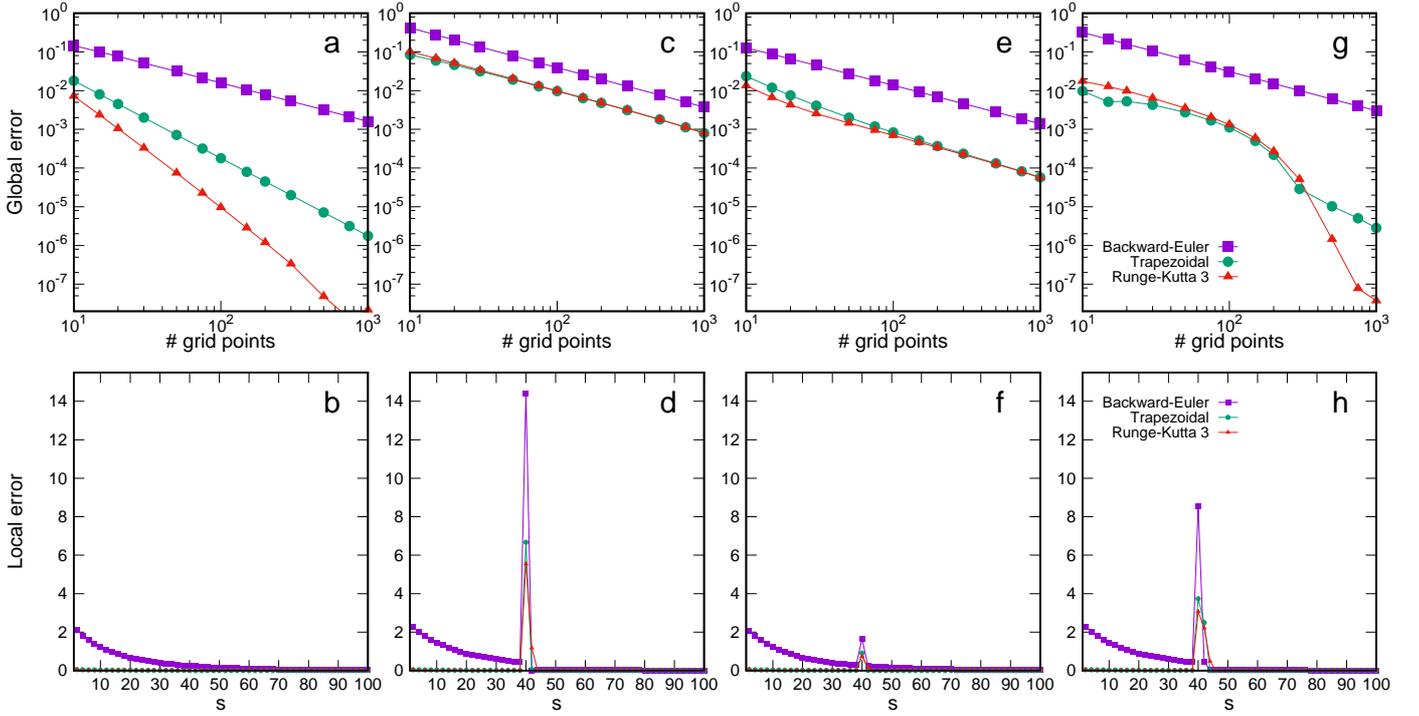}
\caption{Top row: log-log representation
of the global error in the emergent intensity profile as a function of the number of grid-points.
Bottom row: Local error as a function of the spatial coordinate $s\in[0,100]$ for a sampling of 50 equispaced grid points for a single wavelength near the line core.
The four columns correspond to four different atmospheric models, where
the absorption coefficient $\chi_{\nu}$ and emissivity $\epsilon_{\nu}$ are given, respectively,
by Equations~\eqref{case1_k} and~\eqref{case1_e} (first column),
by Equations~\eqref{case2_k} and~\eqref{case2_e} (second column),
by Equations~\eqref{case3_k} and~\eqref{case3_e} (third column), and
by Equations~\eqref{case4_k} and~\eqref{case4_e} (fourth column).
The parameters are:
$c_1=0.01$, $c_2=0.05$, $c_3=0.15$, $k_1=1/25$, $k_2=1/15$, $k_3=5$,
$j_1=2$, $j_2=20$, $s_{d1}=39.985$, and $s_{d2}=39.955$.
This choice of $s_{d1}$ and $s_{d2}$ warrants a monotonic approach of the grid points
to the location of the discontinuity (or sharp gradient) as the grid sampling is refined.
The global and the local error are computed as exposed in Equations~\eqref{error} (considering the Stokes $I$ component only) and~\eqref{local_error}, respectively,
with a reference atmospheric model with $10^4$ grid points.}
\label{fig:convergence_localerror}
\end{figure*}
\begin{table}
\caption{Numerical schemes}
\setlength{\tabcolsep}{5pt}\renewcommand{\arraystretch}{1.5}
\centering
\begin{tabular}{ l c c}
\hline
\hline
\emph{Formal solver}    & \emph{Order}  & \emph{Stability} \\
\hline 
\hline
Backward-Euler          & 1     & $L$-stable    \\
Trapezoidal             & 2     & $A$-stable    \\
DELO-linear             & 2     & $L$-stable\tablefootmark{a}   \\
Pragmatic 2             & 2     & $L$-stable\tablefootmark{a}   \\
Pragmatic 3             & 3     & $L$-stable\tablefootmark{a}   \\
Runge-Kutta 3           & 3     & -             \\
cubic DELO-B\'ezier     & 4     & $L$-stable\tablefootmark{a}\\
cubic Hermitian         & 4     & $A$-stable    \\\hline
\end{tabular}
\tablefoot{
\tablefoottext{a}{$L$-stability is guaranteed for the scalar problem (or diagonal $\mathbf K_{\nu}$ in Equation~\eqref{eq:RTE})}
}
\label{tab:schemes}
\end{table}
\subsection{Scalar case}\label{sec:sec5.1}
This section presents the integration of
Equation~\eqref{eq:scalar_RTE} in four different scenarios,
where the absorption and emission coefficients show, respectively:
\begin{enumerate}[(i)]
 \item a smooth behavior;
 \item a discontinuity in absorption;
 \item a discontinuity in emissivity;
 \item a high gradient in absorption.
\end{enumerate}
Spectral line profiles are synthesized using a modified version of the Octave code used by \citet{janett2017a,janett2017b},
while the spectral line parameters are identical to those described in Appendix C of \citet{janett2017a}.
\subsection*{Case {\rm (i)}: Smooth atmosphere}
In order to verify the predicted order of accuracy of the different numerical schemes,
the case of a smooth atmospheric model is first analyzed. 
Consider Equation~\eqref{eq:scalar_RTE} where
\begin{equation}
\chi_{\nu}(s) = c_1 e^{-k_1s}\,,\\\label{case1_k}
\end{equation}
\begin{equation}
\epsilon_{\nu}(s) = c_2e^{-k_2s} \,.\label{case1_e}
\end{equation}
%
Figure~\ref{fig:convergence_localerror}(a) shows that the various numerical methods effectively achieve their predicted order of convergence.
In this case, the smoothness assumptions are fulfilled and, consequently, the use of high-order methods either allows
to reach higher accuracy or makes the use of coarser spatial grids feasible.
Moreover, Figure~\ref{fig:convergence_localerror}(b) shows that local errors are bigger where absorption and emission gradients are higher.
%
%
\subsection*{Case {\rm (ii)}: Discontinuity in absorption}
The effect of discontinuities in the right-hand side of Equation~\eqref{eq:scalar_RTE} is first analyzed by considering
\begin{equation}
\chi_{\nu}(s) =
\begin{cases}
c_1\left(j_1+e^{-k_1s}\right) & \text{if}\; s<s_{d1}\,, \\ 
c_1 e^{-k_1s} & \text{if}\; s\ge s_{d1}\,, \label{case2_k}
\end{cases}
\end{equation}
\begin{equation}
\epsilon_{\nu}(s) = c_2e^{-k_2s} \,.\label{case2_e}
\end{equation}
A first-order jump of magnitude $j_1$ occurs in the absorption coefficient when the independent variable $s$ reaches the value $s_{d1}$.

Figure~\ref{fig:convergence_localerror}(c) clearly shows the order breakdown due to the first-order discontinuities.
Moreover, Figure~\ref{fig:convergence_localerror}(d) shows that the global error is dominated by the local errors due to the discontinuity.
%
%
\subsection*{Case {\rm (iii)}: Discontinuity in emissivity}
The effect of discontinuities in the right-hand side of Equation~\eqref{eq:scalar_RTE} is analyzed by considering
\begin{equation}
\chi_{\nu}(s) = c_1 e^{-k_1s}\,,\\\label{case3_k}\end{equation}
\begin{equation}
\epsilon_{\nu}(s) 
\begin{cases}
c_2\left(j_2+e^{-k_2s}\right) & \text{if}\; s<s_{d1}\,, \\ 
c_2 e^{-k_2s} & \text{if}\; s\ge s_{d1}\,. \label{case3_e}
\end{cases}
\end{equation}
A first-order jump of magnitude $j_2$ occurs in the emission coefficient when the independent variable $s$ reaches the value $s_{d1}$.

Just like in case (ii), Figure~\ref{fig:convergence_localerror}(e) clearly shows the predicted order breakdown due to the first-order discontinuities.
Moreover, Figure~\ref{fig:convergence_localerror}(f) shows that the global error is dominated by the local errors due to discontinuity.
%
\subsection*{Case {\rm (iv)}: High gradient in absorption}
Atmospheric models can involve rapidly varying physical parameters without an actual mathematical discontinuity. 
The effect of sharp variations in the right-hand side of Equation~\eqref{eq:scalar_RTE} is analyzed with the model
\begin{equation}
\chi_{\nu}(s)= c_1\left(j_1\cdot\frac{1}{2}(1-\tanh(k_3(s-s_{d2}))+e^{-k_1s}\right)\,,\label{case4_k}
\end{equation}
\begin{equation}
\epsilon_{\nu}(s) = c_2e^{-k_2s} \,.\label{case4_e}
\end{equation}
A sharp gradient occurs in the absorption coefficient when the independent variable $s$ reaches the value $s_{d2}$.
For $k_3\rightarrow\infty$ the function $\frac{1}{2}(1+\tanh(k_3(s-s_{d2})))$
tends to a step function with the discontinuity at $s = s_{d2}$.
This case represents the continuous version of case (ii).

For coarse samplings, Figure~\ref{fig:convergence_localerror}(g) clearly shows the order breakdown due to the high gradient.
The global error is dominated by the local errors due to the high gradient and
such errors present only first-order convergence until the sampling starts to resolve the sharp gradient.
In fact, Section~\ref{sec:int_disc_data} explains that
there is no difference between a discontinuity and a sharp gradient
for a set of discrete data which is not able to resolve the local feature.
Figure~\ref{fig:convergence_localerror}(g) shows that, once the sampling resolves the high gradients,
the numerical methods start to converge according their predicted order of accuracy.
%
%
%
%
%
%
%
\begin{figure}
\centering
\includegraphics[width=0.49\textwidth]{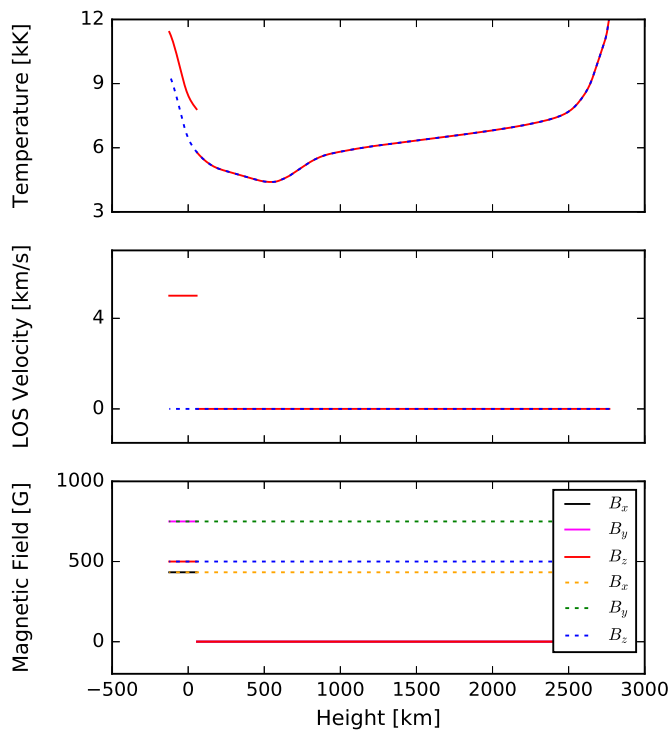}
\caption{Temperature (top), upward directed line of sight velocity (middle), and three components of the magnetic field (bottom),
all as a function of the geometrical height in the atmosphere.
The dashed lines represent the FALC smooth atmospheric model, while solid lines show discontinuities in temperature, velocity, and magnetic field profiles.}
  \label{fig:model_disc}
\end{figure}
\subsection{Polarized case}\label{sec:sec5.2}
This section presents some numerical evidence for the integration of Equation~\eqref{eq:RTE}
in the case of discontinuities in the atmospheric physical parameters.
From these numerical tests, one can still draw conclusions about the scalar problem considering the Stokes $I$ component only.

Stokes profiles of the photospheric Sr~{\sc i} line at 4607.3 {\rm \AA} are synthesized using a modified version of the RH code of \citet{uitenbroek2001}
that allows to switch between different formal solvers and to sequentially perform Stokes-profile syntheses 
with a set of discrete atmospheric models \citep[see][]{janett2018b}.
In these calculations, spectral lines are synthesized under the assumption of LTE conditions and the polarization is produced by the Zeeman effect alone.

Figure~\ref{fig:model_disc} shows the considered atmospheric models. These are
modified versions of the model C of
\citet{fontenla1993}, the so-called FALC model\footnote{An additional magnetic field is provided because FAL models do not include a specific magnetic field.}.
These models present, respectively:
\begin{enumerate}[(i)]
 \item a smooth profile;
 \item a discontinuity in the temperature;
 \item a discontinuity in the line of sight velocity;
 \item a discontinuity in the magnetic field.
\end{enumerate}
These atmospheric models are homogeneously re-sampled in $\log\tau_c$
($\tau_c$ is the continuum optical depth at the wavelength $\lambda=5000$ {\rm \AA})
with different grid-point densities
in the optical depth interval $-8.6\le\log\tau_c\le1.4$, which encompasses 10 decades. 
The Sr~{\sc i} line at 4607.3 {\rm \AA} is sampled with around 500 points equispaced in frequency
in a spectral interval of a few {\rm angstroms} around the core.

\begin{figure*}
\centering
\includegraphics[width=1.\textwidth]{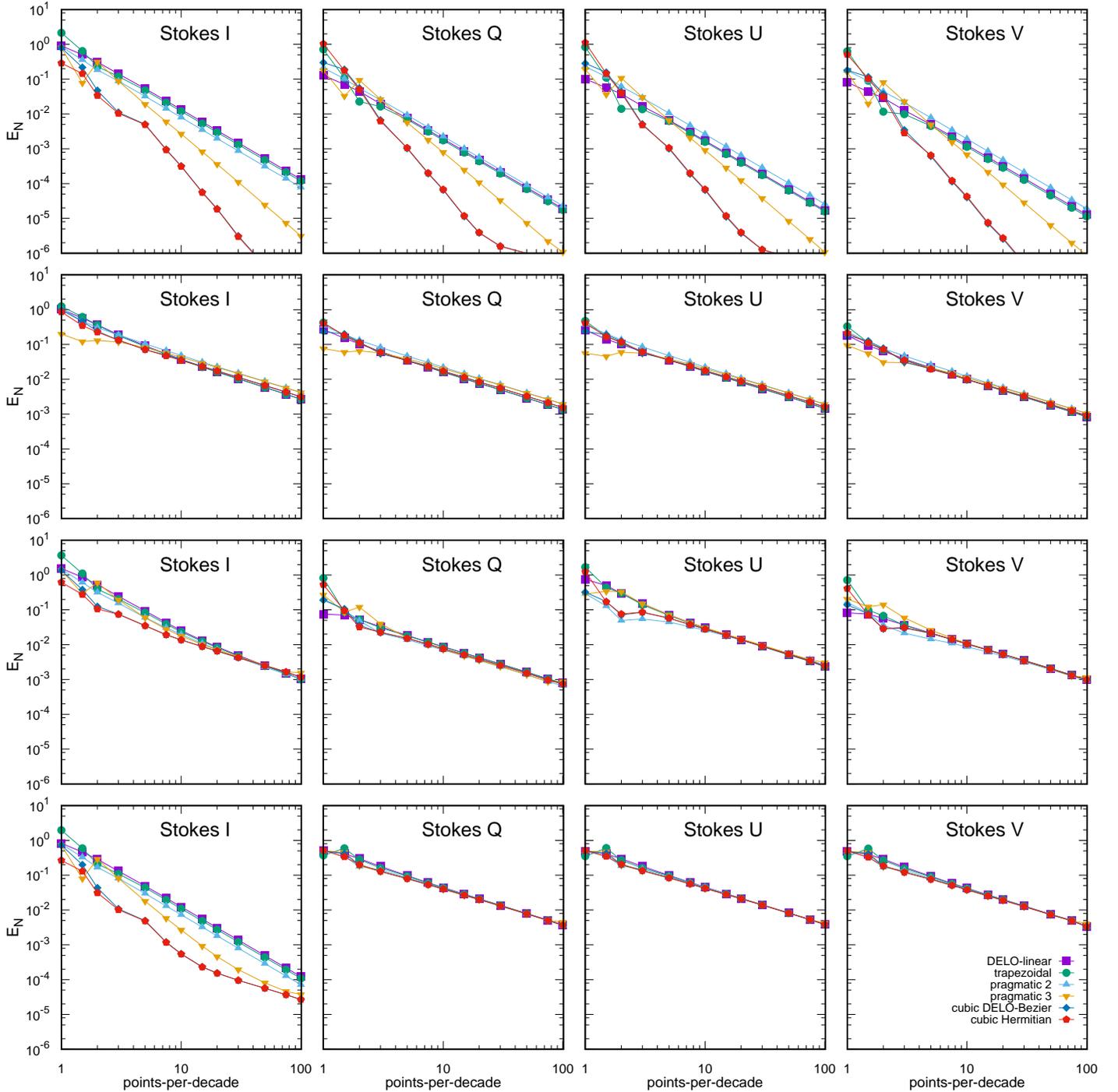}
\caption{The log-log representation of the global error for the Stokes vector components $I,Q,U$ and $V$
as a function of the number of points-per-decade of continuum optical depth for different formal solvers (color coded).
The Sr~{\sc i} line at 4607.3 {\rm \AA} is considered for the smooth FALC atmospheric model (first row),
and for three additional atmospheric models that contain discontinuities in temperature (second row),
in line of sight velocity (third row), and in magnetic field (fourth row).
The atmospheric models are described in Figure~\ref{fig:model_disc}.
The global error is computed as exposed in Equation~\eqref{error}.}
\label{fig:sri_disc}
\end{figure*}

Figure~\ref{fig:sri_disc} gives the log-log representation
of the global error in the emergent Stokes profiles as a function of the number of points-per-decade of continuum optical depth,
i.e., the number of grid points that sample a variation of one
order of magnitude in the optical depth at the wavelength $\lambda=5000$ {\rm \AA}.
These error curves allow for different general considerations.

In the pre-asymptotic regime (below 2 points-per-decade),
the comparison between the rows of Figure~\ref{fig:sri_disc} does not reveal essential differences in the error curves.
In this regime, the accuracy of numerical schemes strongly depends on the specific sampling.

Asymptotic convergence rates become relevant above 3 points-per-decade.
The first row of Figure~\ref{fig:sri_disc} shows that the different numerical methods
effectively achieve the predicted (second, third, and fourth) order of accuracy.
In this case, the smoothness assumptions are fulfilled,
allowing the asymptotic order of accuracy of the numerical solution and
high-order schemes out-perform low-order methods.

By contrast, the second, third, and fourth rows of Figure~\ref{fig:sri_disc} clearly exhibit
the order breakdown due to the discontinuity.
In fact, all the numerical methods drop to first-order convergence,
making the application of high-order schemes pointless.
This attests to the fact that discontinuities in the atmospheric physical parameters effectively
induce first-order discontinuities in Equation~\eqref{eq:RTE} and
affect the convergence (and the accuracy) of numerical solutions.
In the fourth row, the error curve of Stokes $I$ seems to be only mildly affected by the discontinuity.
This is because the magnetic field has a meager impact on Stokes $I$ with respect to $Q$, $U$, and $V$.

Dropped-to-first-order methods require among 30-40 points-per-decade to achieve an accuracy of $E_i\le10^{-2}$, for $i=1,2,3,4$ (see Equation~\ref{error}),
whereas they require prohibitive numerical grids to produce $E_i\le10^{-3}$ accurate spectra.
This demonstrates the need for high-order well-behaved
formal solvers able to face discontinuities in the radiative transfer equation.
\section{Numerical methods for discontinuous ODEs}\label{sec:sec4}
The need for efficient and accurate numerical methods for treating discontinuous differential systems has been widely recognized.
Most of the literature on the topic is focused on adaptive stepsize numerical
methods, that is, numerical schemes that dynamically choose the length of steps along the integration. 
In this way, the algorithm maintains the desired accuracy of the approximation by decreasing the step size when large derivatives
(or singularities)
appear and improves efficiency by increasing the step size whenever the smoothness of the system allows it.
In real-life radiative transfer problems, the sampling of the atmospheric model is given.
This implies that the step size is fixed \textit{a priori} and thus the adaptive step-size technique cannot be applied.
In principle, one might circumvent this problem by providing the absorption and emission quantities
at intermediate grid-points through interpolations.
However, Section~\ref{sec:sec31} points out that interpolations
suffer from local oscillations around discontinuities.

When facing a discontinuous ODE, \citet{gear1984} distinguish among four main stages in a numerical procedure:
\begin{enumerate}
 \item detecting the discontinuity,
 \item locating the discontinuity,
 \item crossing the discontinuity,
 \item restarting after the discontinuity.
\end{enumerate}
Moreover, \citet{dieci2012} classified the different numerical schemes in three main categories:
\emph{time stepping} methods, \emph{event driven} methods, and \emph{regularization} of the differential system. 
This section briefly reviews the numerical schemes for the treatment of discontinuous ODEs,
investigating their applicability to Equations~\eqref{eq:scalar_RTE} and~\eqref{eq:RTE} in discontinuous media.
\subsection{Time stepping methods}
Time stepping methods usually rely on local error estimators to detect and locate discontinuities:
a large value of the local error estimate indicates the presence of a discontinuity and
different strategies to estimate local errors are available \citep{shampine1971,butcher1993}.
Alternatively, \citet{calvo2003} exposed the possibility to detect discontinuities by using the so-called pairs of embedded forms
or by monitoring the size of an estimate of the local defect.
Some discontinuities might avoid detection. These would probably have negligible effects
on the local truncation error and they can consequently be ignored.
In contrast, for some smooth but rapidly varying problems, one may detect discontinuities that
mathematically do not exist. 
In this case, 
it is often reasonable to
assume that a discontinuity is present and to act accordingly \citep{gear1984}.

In radiative transfer problems, one can
determine at each step whether or not the numerical
solution is crossing a discontinuity.
This detection should be computationally inexpensive and must fit well into the code, that is, it should be based on quantities that the code computes at each step.
Alternatively, one may locate discontinuities using the discrete values of absorption and emission coefficients and
by detecting the discontinuities in the discrete data before the integration
or during the integration.
This approach involves data analysis along the entire ray path and 
has the advantage that the treatment
of the discontinuity is performed in a separate preliminary step.
A more in-depth study is required to support this strategy,
but some suitable techniques are already available \citep{gelb1999,gelb2002,mallat2006,oeffner2013}. 

In real-life radiative transfer problems, the location of a discontinuity is, in some sense, harder to decipher than commonly assumed.
In discrete data,
it is usually impossible to exactly locate the discontinuity, because there is a lack of knowledge and the jump can occur anywhere
inside the interval.

After having located a discontinuity, time stepping methods adapt time steps to ensure that the local error is kept below a specified tolerance $\epsilon$.
This automatically enforces accuracy in spite of the decreased smoothness of the problem.
Accordingly, \citet{dieci2012} defined the passing stepsize $\Delta \tilde{t}$ as
\begin{equation*}
\Delta\tilde{t}=\frac{\epsilon}{K_1}\,,
\end{equation*}
where $K_1$ is the first-order jump in $f$ defined in Section~\ref{sec:subsec2.2}.
However, radiative transfer problems usually assume \textit{a priori} fixed step sizes and the adaptive techniques are not suitable.


After having crossed a discontinuity in $[t_{k-1},t_k]$, the numerical methods start up the usual integration.
However, one must be careful about the previously obtained information,
because back values of $y$ and $f$ (i.e., $y_{k-1},f_{k-1},y_{k-2},f_{k-2},\ldots$) may not correspond to those
of smooth functions. In case of a discontinuity of order $q$, they differ by terms of the order $\mathcal{O}(h^q)$ and $\mathcal{O}(h^{q-1})$, respectively.
The use of back values in the numerical integration give rise to terms in the local truncation error of order $\mathcal{O}(h^q)$, and
clearly, it is always safe to use a numerical method that does not use them (e.g., the trapezoidal method and DELO-linear).
\subsection{Event-driven methods}
In many problems, discontinuities do not occur completely unexpectedly.
The so-called event-driven methods assume that there is an \emph{event function}, i.e., a function that changes sign at the location of the discontinuity \citep{dieci2012}.
\citet{mao2002} pointed out that most discontinuity-detection algorithms are based on the root-finding of event functions.
In fact, when the numerical solution reaches a root of the event function the discontinuity is detected and located.
Once the location of the discontinuity is precisely known, then the numerical method includes it as an extra grid point and
the algorithm accordingly adjusts the stepping and restarts at that point.
In doing so, the local smoothness assumptions are fulfilled,
maintaining the asymptotic correctness of the numerical solution and avoiding the order breakdown \citep{mannshardt1978}.

Unfortunately, no event function is known in the radiative transfer problems and event-driven methods are consequently not suitable in this context.
\subsection{Regularization of the differential system}
An alternative strategy is to regularize (or smoothen) the differential system, removing the mathematical discontinuities. 
Undoubtedly, this leads to simplifications in the theory on ODE\footnote{The Picard-Lindel\"of Theorem ensures the existence and uniqueness of a solution
to the IVP.}. A more technical and thorough investigation of this topic can be found in \citet{llibre1997} and \citet{llibre2007}.
However, due to the large derivatives that replace the structural discontinuities, the regularized system becomes quite stiff and small time steps are then required.
Moreover, regularization can lead to changes in the dynamics of the original nonsmooth system.
For these reasons, the regularization of the differential system strategy is not particularly adequate for radiative transfer problems.
%
%
%
\section{Interpolations for discontinuous discrete data}\label{sec:int_disc_data}
Different approximations can be used to recover the various intermittent quantities
involved in radiative transfer problems at off-grid points.
The goal is to obtain high-order accuracy in smooth regions,
avoiding spurious effects around sharp gradients and discontinuities.
The standard strategy is to use a locally monotonic\footnote{In practice,
one requires interpolants of monotonic data to themselves be monotonic.} interpolation
for each direction ($x$, $y$, $z$ and the
short-characteristic direction in Cartesian grids).

This section shortly reviews different interpolation (or reconstruction) techniques for scattered quantities,
providing some illustrative applications to the radiative transfer problem conducted in the past.
In particular: cubic Hermite splines, B\'ezier curves, piecewise rational polynomials, slope limiters,
and high-order (weighted) essentially non-oscillatory approximations.
%
%
\subsection{Cubic Hermite splines}\label{subsec:cubic_hermite}
Given a set of points $\{x_i\}$, the Hermite interpolation $H$ does not only match a set of function values $\{y_i\}$,
but also its derivatives \citep[see][Section 4]{janett2017b}.
%
For notational simplicity one defines the normalized variable $t \in [0,1]$ as
\begin{equation*}
t=\frac{x-x_i}{\Delta x_i}\,,\text{ for } x \in [x_i,x_{i+1}]\,.
\end{equation*}
The cubic Hermite interpolation, approximating the function $y(t)$ inside the interval $t \in [0,1]$, reads
\begin{equation}
\begin{split}
y(t)&\approx y_i\cdot(1-3t^2+2t^3) + y_i'\cdot \Delta x_i(t-2t^2+t^3)\\
&+ y_{i+1}\cdot (3t^2-2t^3) + y_{i+1}'\cdot \Delta x_i(-t^2+t^3)\,.
\label{hermite_cubic}
\end{split}
\end{equation}
It is known that the cubic Hermite interpolant is fourth-order accurate if the derivatives are at least third-order, and
third-order accurate if the derivatives are second-order, and so on \citep{dougherty1989}.
Therefore, assuming derivatives that are at least third-order accurate, the error scales as $\mathcal{O}(h^4)$,
indicating that the cubic Hermitian interpolation is fourth-order accurate.
Monotonicity can be ensured by a suitable choice of the derivatives.
The main weakness of this approach is that it necessarily degenerates
to a linear interpolation near smooth extrema \citep{shu1998}.

\citet{fritsch1980} proposed a two-pass algorithm to recover such derivatives.
Later on, 
\citet{fritsch1984} and \citet{steffen1990}
added alternative formulas to recover second-order accurate first derivatives
\footnote{The formula by \citet{fritsch1984} is
second-order accurate on uniform grids only and
it drops to first-order on nonuniform grids.}.
\citet{auer2003} suggested the use of monotonic Hermite interpolants in radiative transfer problems
and \citet{ibgui2013} used monotonic cubic Hermite polynomials \citep[version of][]{fritsch1984} in the IRIS code.
\subsection{B\'ezier curves}\label{subsec:bezier_curves}
These well-known interpolations make use of the so-called control points (or weights) to suppress spurious extrema.
A B{\'e}zier curve of degree $q$ applied to a set of function values $\{y_i\}$ at positions $\{x_i\}$ 
inside the interval $t \in[0,1]$ can be defined as
\begin{equation*}
B_q(t)=\sum_{n=0}^{q} C_n B_{n,q}(t)\,,
\end{equation*}
where $C_n$ are the control points, and the Bernstein polynomials $B_{n,q}(t)$ are given by
\begin{equation*}
B_{n,q}(t)=\binom{q}{n}\cdot t^n\left(1-t\right)^{q-n}\,.
\end{equation*}
The first and the last control points define the start and end points of the B{\'e}zier curve in the interval, that is,
\begin{equation*}
C_0 = y_i\,,\text{ and }C_q = y_{i+1}\,.
\end{equation*}
All the remaining points, conventionally referred to as weights, are usually used to shape the curve.
Moreover, a B{\'e}zier curve always lies in the convex hull of the control points, that is,
in the smallest set that contains the line segment joining every pair of control points \citep[see][Section 5]{janett2017b}.
One can avoid the creation of new extrema by adjusting the weights.
However, the high-order accuracy of B{\'e}zier interpolations is
achieved by forcing the B{\'e}zier interpolants to be
identical to the corresponding degree Hermite interpolants but, in this case, monotonicity is not guaranteed.

\citet{auer2003} suggested the use of B{\'e}zier curves in radiative transfer problems,
because of their suitability for preventing spurious behavior near rapid variations in the absorption and emission coefficients.
Moreover, \citet{stepan+trujillo_bueno2013} used B{\'e}zier interpolations in the PORTA code, whereas
\citet{delacruz_rodriguez+piskunov2013} made use of them to construct the quadratic and cubic DELO-B{\'e}zier formal solvers.
\subsection{Piecewise rational polynomials}\label{subsec:pw_rational}
As an alternative to the standard use of polynomials for the interpolation of monotonic data,
\citet{gregory1982} and~\citet{delbourgo1983} opted for the application of piecewise
rational quadratic functions. 
\citet{delbourgo1985} constructed a monotone, piecewise rational cubic interpolation,
that includes the rational quadratic function as a special case.

A piecewise rational cubic function $R(t)$, approximating the function $y(t)$ with $t\in[0,1]$, reads
\begin{equation}
R(t)=\frac{P(t)}{Q(t)}\,,
\label{rat_funct}
\end{equation}
where
\begin{equation}
\begin{split}
P(t)&= y_{i+1}\cdot t^3 + (r_i y_{i+1}-\Delta x_i y_{i+1}')\cdot t^2(1-t)\\
& +(r_i y_{i}+\Delta x_i y_{i}')\cdot t(1-t)^2+y_{i}\cdot (1-t)^3\,,
\label{P_cubic}
\end{split}
\end{equation}
and
\begin{equation}
Q(t)= 1+(r_i-3)t(1-t)\,.
\label{Q_cubic}
\end{equation}
The condition $r_i>-1$ ensures a strictly positive denominator in the rational cubic,
while $r_i=3$ reduces the rational cubic to the standard cubic Hermite polynomial given in Equation~\eqref{hermite_cubic}.
Here, the parameter $r_i$ is chosen to ensure that the interpolant preserves monotonicity.
In particular, if
\begin{equation*}
r_i= 1+\Delta x_i\frac{y_i'+y_{i+1}'}{y_{i+1}-y_i}\,,
\end{equation*}
then the rational cubic defined by Equation~\eqref{rat_funct} reduces to the rational quadratic form
{\small
\begin{equation}
R(t)= \frac{y_{i+1}t^2+\Delta x_i t(1-t)(y_{i+1}y_i'+y_iy_{i+1}')/(y_{i+1}-y_i)+y_i(1-t)^2}{t^2+\Delta x_i(y_i'+y_{i+1}')t(1-t)/(y_{i+1}-y_i)+(1-t)^2}\,,
\label{rat_funct2}
\end{equation}}
which yields a monotonic interpolant.

In most applications, the derivatives $y_i'$ and $y_{i+1}'$ are not known in advance and hence must be numerically determined.
\citet{delbourgo1985} proposed different approximations for the derivative parameters
and showed that an error $\mathcal{O}(h^4)$ can be expected when $\mathcal{O}(h^3)$ derivative information is given at the data points.
Finally, the rational quadratic given by Equation~\eqref{rat_funct2} can be used to construct a $C^2$ rational spline
which interpolates strictly monotonic data \citep{delbourgo1985}.
\subsection{Slope limiters}\label{subsec:slope_limiters}
An alternative method to avoid spurious oscillations near discontinuities is to reduce the order of accuracy of the interpolation,
for example using a linear rather than a quadratic interpolant near the discontinuity,
or by reducing the slope of reconstructions applying limiters.
In fact, large gradients and discontinuities produce local extrema in high-order reconstructions, inducing a lack of monotonicity (overshoots), demonstrating the necessity to limit the spatial derivatives inside each cell to physically meaningful values.
This is already guaranteed for the Godunov (constant) scheme, but some limitations are required for the Van Leer (linear) and the parabolic reconstructions.
In computational fluid dynamics, this problem is solved by the so-called slope limiters.
There exists a variety of slope limiters 
and broad literature attests to the importance of this topic \citep[e.g.,][]{colella+woodward1984,sweby1984,berger2005,colella2008,velechovsky2013}.
One disadvantage of this approach is that it necessarily degenerates to a linear interpolation near smooth extrema \citep{shu1998}.

\citet{steiner2016} introduced the use of limiters for the interpolations and reconstructions of the
coefficients involved in the polarized radiative transfer equation, purposely taking discontinuities
into account.
\subsection{Essentially non-oscillatory and weighted essentially non-oscillatory approximations}\label{eno_weno}
%
Provided the smoothness of the function inside the interval,
it is known that the wider the stencil, i.e., the set of grid points considered, the higher the order of accuracy of the interpolation.
The most common interpolations (see Sections~\ref{subsec:cubic_hermite}--\ref{subsec:pw_rational}) are based on fixed stencils.
However, fixed stencil interpolations of second or higher order are oscillatory in the presence of a discontinuity.

Essentially non-oscillatory (ENO) methods are based on a nonlinear adaptive procedure that automatically chooses the stencil that leads
to the locally smoothest interpolant, avoiding crossing discontinuities.
A broad range of literature on the topic is available \citep{liu1994,shu1998,shu2009}.
The classical ENO scheme chooses, among several candidates, the smoothest stencil to work with and discards the rest.
This stencil is then used to construct a polynomial interpolation.
Otherwise, weighted ENO (WENO) schemes use weighted linear combinations of smaller-stencil polynomials
to obtain higher-order accuracy.
WENO approximations guarantee a non-oscillatory
result since the contribution from any stencil containing the discontinuity has an
essentially zero weight.

Both ENO and WENO schemes are especially suitable for problems containing both discontinuities and
smooth structures.
Moreover, they do not necessarily degenerate
to a linear interpolation near smooth extrema.

As an explicit example, an illustrative third-order accurate ENO interpolation is presented in the following.
%

To obtain a third-order-accurate interpolation of the function $y(x)$, one needs a three-point stencil.
Thus in cell $[x_i,x_{i+1}]$, one starts with the two-point stencil $S_2=\{x_i,x_{i+1}\}$.
Subsequently, one has two choices to expand the stencil: by adding either the left neighbor $x_{i-1}$
or the right neighbor $x_{i+2}$.
The selection criteria is to compare the local smoothness of the function to be interpolated,
measured in terms of divided differences \citep{shu1998}.
For instance, one takes the decision by comparing the absolute values of the two relevant divided differences
$y [x_{i-1},x_i,x_{i+1}]$ and $y [x_i,x_{i+1},x_{i+2}]$.
A smaller one implies that the function is smoother in that stencil.
Therefore, if
$$|y [x_{i-1},x_i,x_{i+1}]| < |y [x_i,x_{i+1},x_{i+2}]|\,,$$
the three-point stencil is taken as
\begin{equation*}
S_3 = \{x_{i-1},x_i,x_{i+1}\}\,, 
\end{equation*}
otherwise,
\begin{equation*}
S_3 = \{x_i,x_{i+1},x_{i+2}\}\,. 
\end{equation*}
Once the interpolation stencil is determined, one proceeds with a standard (Lagrange) polynomial
interpolation.
Clearly, one can continue this procedure to add more grid points to the stencil,
constructing higher-order ENO approximations.

A more thorough investigation on the possible applications of ENO and WENO techniques
in radiative transfer problems is in progress.
\section{Conclusions}\label{sec:sec6}
This paper identifies and exposes the relevant problems appearing in the numerical treatment of the radiative transfer equation in discontinuous media.
The main aim is not to provide the ultimate numerical method able to face discontinuous problems,
but to better understand the specific situations where discontinuity issues appear.

The first part pays 
particular attention to assumptions and limitations of the standard convergence analysis
of numerical schemes for ODEs and interpolations.
In fact, standard numerical methods (and interpolations) rely on smoothness assumptions on the functions
to be integrated (interpolated)
and may perform very inefficiently in the presence of discontinuities,
which may drastically increase local errors, reducing the accuracy of the solution
and thwarting high-order convergence.

The numerical tests
(performed for both the scalar and the polarized case) corroborate analytical predictions:
discontinuities in absorption and emission coefficients effectively
affect the convergence and, consequently, the accuracy of numerical solutions,
inducing the order breakdown where all the numerical methods drop to first-order convergence.
Moreover, local errors are higher around high-gradients
and unresolved high-gradients behave like discontinuities.
It is also shown that discontinuities in the atmospheric physical parameters effectively
induce first-order discontinuities in the radiative transfer equation,
making the application of high-order schemes pointless.
Unfortunately, first-order convergence is not sufficiently accurate in many situations.
In the presence of a discontinuity, the order breakdown in the formal solution is inevitable and
the only practical solution is to increase resolution in the atmospheric model.
Discontinuities do not cause numerical instability (in the sense of magnification of errors)
in the formal solution, but they just decrease accuracy.


The final part summarizes the existing numerical methods for the treatment of discontinuous ODEs
and the interpolation techniques for discontinuous discrete data.
Standard numerical schemes for discontinuous ODE are shown to be unsuitable for common RTE problems.
By contrast, many suitable interpolation techniques are already available. In particular, 
Essentially Non-Oscillatory (ENO) and Weighted ENO (WENO) techniques
are high-order robust approximations that
resolve discontinuities
in an accurate and non-oscillatory fashion.
An investigation into
their possible applications
to radiative transfer problems is in progress.
%
%
\begin{acknowledgements}
The financial support by the Swiss National Science Foundation (SNSF) through grant ID 200021\_159206 is gratefully acknowledged.
Special thanks are extended to F. Calvo and C. Bertoni for particularly enriching discussions, and
to E. Alsina Ballester, L. Belluzzi, and O. Steiner
for reading and commenting on previous versions of the paper.
\end{acknowledgements}

\appendix

\section{Error calculation}\label{appendix:A}
Denoting with $\mathbf I^{\rm ref}(\nu)$ and $\mathbf I^{\rm num}(\nu)$ the reference and the numerically computed emergent Stokes vectors, respectively, at the frequency $\nu$, the global error for each Stokes vector component is computed as 
\begin{equation}
E_i=\frac{\displaystyle \max_{\nu}| I_i^{\rm ref}(\nu)- I_i^{\rm num}(\nu)|}{{\displaystyle \max_{\nu}} \; I_i^{\rm ref}(\nu)-{\displaystyle\min_{\nu}} \; I_i^{\rm ref}(\nu)}\,,
\text{ for }i=1,2,3,4\,,
\label{error}
\end{equation}
where $i=1,2,3,4$ indicates the Stokes parameters $(I,Q,U,V)$. The error is given by the maximal discrepancy between the reference and the simulated Stokes parameter over the spectral interval considered, normalized by the maximal amplitude in the reference profile. The reference emergent Stokes profile is calculated with the cubic Hermitian method (and cross-checked with the cubic DELO-B\'ezier method) using a hyperfine grid sampling with $10^3$ points-per-decade of continuum optical depth.

Denoting with $I_k^{\rm ref}(\nu)$ and $I_k^{\rm num}(\nu)$ the reference and the numerically computed specific intensities at the frequency $\nu$ and at the grid location $s_k$,
the local error is computed as
\begin{equation}
L_k(\nu)= | I_k^{\rm ref}(\nu)- I_k^{\rm num}(\nu)|\,,
\text{ providing }I_{k-1}^{\rm num}(\nu)=I_{k-1}^{\rm ref}(\nu)\,.
\label{local_error}
\end{equation}

\bibliographystyle{aa}
\bibliography{bibfile_disc}

\appendix
\end{document}